\documentclass[fleqn,usenatbib]{mnras}

\usepackage[T1]{fontenc}

\DeclareRobustCommand{\VAN}[3]{#2}
\let\VANthebibliography\thebibliography
\def\thebibliography{\DeclareRobustCommand{\VAN}[3]{##3}\VANthebibliography}

\usepackage{graphicx}	
\usepackage{amsmath}	

\renewcommand{\vec}[1]{ {\mathbf #1} }

\newcommand{\Fig}{{Figure}}

\title[Model of Magnetic Flux Rope Eruption]
{A Model of Solar Magnetic Flux Rope Eruption Initiated Primarily by Magnetic Reconnection}

\author[Liu et al.]{Qingjun Liu,$^{1}$ Chaowei Jiang,$^{1,2}$\thanks{E-mail:
    chaowei@hit.edu.cn (CWJ)} Xinkai Bian,$^{1}$ Xueshang Feng,$^{1,2}$
    Pingbing Zuo,$^{1,2}$ and Yi Wang$^{1,2}$
  \\
  $^{1}$Shenzhen Key Laboratory of Numerical Prediction for Space
  Storm, Institute of Space Science and Applied Technology,\\ Harbin
  Institute of Technology, Shenzhen 518055, China\\
  $^{2}$Key Laboratory of Solar Activity and Space Weather, National
  Space Science Center, \\Chinese Academy of Sciences, Beijing 100190,
  China}

\date{Accepted XXX. Received YYY; in original form ZZZ}

\pubyear{2023}

\begin{document}
\label{firstpage}
\pagerange{\pageref{firstpage}--\pageref{lastpage}}
\maketitle

\begin{abstract}
  There is a heated debate regarding the specific roles played by ideal magnetohydrodynamic (MHD) instability and magnetic reconnection in the causes of solar eruptions. In the context with a pre-existing magnetic flux rope (MFR) before an eruption, it is widely believed that an ideal MHD instability, in particular, the torus instability, is responsible for triggering and driving the eruption, while reconnection, as invoked in the wake of the erupting MFR, plays a secondary role. Here we present a new numerical MHD model in which the eruption of a pre-existing MFR is primarily triggered and driven by reconnection. In this model, a stable MFR embedded in a strapping field is set as the initial condition. A surface converging flow is then applied at the lower boundary, pushing magnetic flux towards to the main polarity inversion line. It drives a quasi-static evolution of the system, during which a current layer is built up below the MFR with decreasing thickness. Once reconnection starts in the current sheet, the eruption commences, which indicates that the reconnection plays a determining role in triggers the eruption. By further analyzing the works done by in the magnetic flux of the pre-existing MFR and the newly reconnected flux during the acceleration stage of the eruption, we find that the latter plays a major role in driving the eruption. Such a model may explain observed eruptions in which the pre-eruption MFR has not reached the conditions for ideal instability.
\end{abstract}




\section{Introduction}
\label{sec:intro}
The magnetic fields in solar corona are line-tied at the solar surface
(namely, the photosphere), and are often highly stressed as driven by
continuous slow motions at the photosphere, for example, the
large-scale shearing, rotational, and converging motions associated
with magnetic flux emergence, and also the small-scale super-granular
and granular convections. As such, magnetic energy is continually
injected into the corona by the Poynting flux associated with the photospheric motions,
and the free energy of the coronal magnetic field often
builds up over several days during which the system evolves in a
quasi-static way. At a critical point, there is a catastrophic
disruption of this quasi-static evolution phase and magnetic energy
releases rapidly, resulting in solar eruptions such as solar flares and
coronal mass ejections (CMEs) with large clouds of magnetized plasma
unleashed into the interplanetary space. Although solar eruptions have
been observed for over a century, it remains unclear how solar
eruptions are initiated, for which the key questions are how the
eruption is triggered and what drives the acceleration of
eruption. Current available theories can be classified into two
categories, one is based on resistive magnetohydrodynamic (MHD)
process (i.e., magnetic reconnection) and the other is based on ideal
MHD process (i.e., ideal instabilities). There is a long-standing
controversy on the specific roles of these two mechanisms in
triggering and driving eruption, largely due to the difficulty in
measuring directly the coronal magnetic fields~\citep{yangResearchProgressCoronal2022}.
		
The different mechanisms often assume different topologies of coronal
magnetic field prior to eruption~\citep{2020SSRv..216..131P}. The
reconnection-based mechanisms are developed based on sheared magnetic
arcade, a single or multiple arcade configuration, for instance, the
classic breakout model~\citep{antiochosModelSolarCoronal1999} and
tether-cutting model~\citep{2001ApJ...552..833M}. The magnetic
breakout model requires the initial magnetic configuration of a
quadrupolar topology that consists of an internal arcade and an
external arcade with inverse directions to each other. As the internal
arcade experiences shear motions along the polarity inversion line
(PIL), it expands upward, leading to the formation of a current sheet
and magnetic reconnection at the interface between the internal and
external arcades. This reconnection continuously reduces the magnetic
tension force that stabilizing the internal arcade, which can then
expand outward (driven by its magnetic pressure force) and enhance the
reconnection, therefore establishing a positive feedback loop. It
results in the drastic eruption when the outward magnetic pressure
force of the inner arcade dominates over the inward magnetic tension
force. The tether-cutting model proposes that eruption can also be
initiated with in a single sheared arcade. When the two J-shaped
sheared magnetic loops gradually approach each other, an internal
current sheet forms above the PIL, and then reconnection occurs in the
current sheet. This reconnection plays a crucial role in reducing the
downward tension of the magnetic arcades by effectively ``cutting the
tethers''. Similar to the breakout model, this process, as going on,
can eventually lead to eruption when the upward magnetic pressure
force can no longer be restrained by the magnetic tension
force. Recently, with a series of high-resolution MHD simulations,
\citet{2021NatAs...5.1126J} demonstrated that once an internal current
sheet is formed in a strongly-sheared arcade, fast reconnection can immediately
trigger an eruption, and the central engine for driving the
eruption comes from the slingshot effect of the reconnection, i.e., the upward magnetic tension force of the newly reconnected
field lines. Therefore, reconnection plays a fundamental role in both
triggering and driving the eruption from a sheared arcade magnetic
configuration.
		
The ideal MHD instability mechanisms are developed based on magnetic
flux rope (MFR), which is a coherent group of magnetic field lines
winding about a common axis by over one full
turn~\citep{2017ScChD..60.1383C, 2020RAA....20..165L}. In this
framework, two primary models for initiation of eruption have been
proposed, including the kink
instability~\citep{hoodKinkInstabilitySolar1979,
  2005ApJ...630L..97T,2007AN....328..743T,2022A&A...667A..89G} and the
torus instability~\citep{kliemTorusInstability2006, 2010ApJ...708..314A,2007ApJ...668.1232F}. The
kink instability arises when the degree of helical twisting of field lines around
the rope's axis exceeds a critical threshold. This
threshold is typically in the range of $1.25 \sim 2.5$ turns of
winding of the field lines~\citep{mikicDynamicalEvolutionTwisted1990,
  fanEmergenceTwistedMagnetic2003, torokEvolutionTwistingCoronal2003,
  2005ApJ...630L..97T,
  torokIdealKinkInstability2004}. As a result of this instability, an
eruption is triggered with rapid writhe of the MFR axis (thus its
center part rises impulsively), while the field lines around the axis
unwinds. This process is accompanied by the release of the stored
magnetic energy within the flux rope. However, the kink instability
saturates quickly and cannot drive a full eruption, which needs the
onset of the other instability, i.e., the torus instability. It
describes the loss of force balance between the MFR and the overlying
field (or the external field).\footnote{The overlying field is also
  referred to as the background field, the external field, or the
  strapping field, which are all used interchangeably in the context
  of describing torus instability.}  The upward forces, consisting of magnetic
pressure and hoop force of the MFR, counteract the downward
confinement forces of magnetic tension of the overlying field. This
instability arises when the external strapping field decay
sufficiently rapidly with height, causing the upward forces on the MFR
to surpass the downward forces during its rise. As a result, the MFR
undergoes a continuous uplift, as driven by the net upward force. To
quantify the rate of decay of the external strapping field, a decay
index is introduced as $n(z) = -\partial\ln B_{e}/\partial\ln z $
\citep{1978mit..book.....B}, where $B_e$ is the external field and $z$
the height or radial distance from solar surface. Theoretical analyses
predict that the critical threshold for the occurrence of torus
instability is approximately $n_c=1.5$~\citep[but still with a wide
range of $1.1 \sim 1.7$ depending on the specific geomoetry of the
MFR, e.g.,][]{kliemTorusInstability2006,
  2007AN....328..743T, 2007ApJ...668.1232F,
  2010ApJ...708..314A,
  demoulinCRITERIAFLUXROPE2010, fanERUPTIONCORONALFLUX2010,
  zuccarelloCRITICALDECAYINDEX2015}, and when the MFR axis reaches the
domain with $n(z) > n_c$, it can erupt successfully.
		
Although a pre-existing MFR can be activated by the aforementioned
ideal instabilities, reconnection is unavoidably associated with its
subsequent eruption. This is due to the stretching of the overlying
magnetic field by the erupting MFR, leading to the dynamic formation
of a current sheet beneath it. Then reconnection is triggered at the
current sheet, producing a flare with the eruption of the MFR. It is
generally believed that in this process, the eruption is driven mainly
by the ideal instability (in particular, the torus instability), while
the flare reconnection is only a by-product of the eruption and plays
a secondary role in driving the eruption by transferring the overlying
flux into the MFR. In this paper we present a three-dimensional (3D) numerical MHD model
that is at variance with this conventional view. We demonstrate a case
in which the eruption of a pre-existing MFR is primarily triggered and
driven by reconnection. The simulation is started from a stable MFR
embedded in a strapping field, and then a converging flow is applied
at the bottom surface (i.e., the photosphere) to the strapping
field. The converging motion drives a quasi-static evolution of the
system, during which the decay index at apex of the MFR axis slowly
increases and meanwhile a current layer is built up below the MFR with
decreasing thickness. At a critical point when the decay index is very
close to the canonical threshold of $n_c = 1.5$, the MFR erupts,
suggesting that the torus instability sets in. Interestingly, the
start of the eruption also coincides with the onset of reconnection in the
current sheet as formed from thinning of the current layer during the
quasi-static evolution phase. This is confirmed by experiments of
different grid resolutions, from which it is found that with higher
resolution the eruption onset time is delayed, since a thinner current
sheet can be supported and thus the reconnection is also delayed. This
indicates that the reconnection plays a determining role in triggers
the eruption. By further analyzing the works done by in the magnetic
flux of the pre-existing MFR and the newly reconnected flux during the
acceleration stage of the eruption, we find that the latter plays a
major role in driving the eruption.

\section{Numerical model}
\label{model}
We used the conservation element and solution element (CESE) method
implemented on an adaptive mesh refinement (AMR) grid with parallel
computing, namely, the AMR--CESE--MHD code~\citep{2010SoPh..267..463J}
to solve the full MHD equations in a 3D Cartesian geometry. The MHD
equations are given as
\begin{eqnarray}
  \frac{\partial \rho}{\partial t}+\nabla \cdot(\rho \vec v) =  0, \nonumber \\
  \rho\frac{D\vec v}{D t}=-\nabla p+\vec J\times \vec B+\nabla \cdot(\nu \rho\nabla \vec v),\nonumber \\
  \frac{\partial \vec B}{\partial t}=\nabla \times (\vec v \times \vec B - \eta\mu_0 \vec J), \nonumber \\
  \frac{\partial T}{ \partial t} +\nabla \cdot(T \vec v)=(2-\gamma)T\nabla\cdot \vec v,
\end{eqnarray}
where $\vec J=\nabla \times \vec B/\mu_0$ and $\mu_0$ is the magnetic
permeability in a vacuum. The momentum equation includes a term for
kinetic viscosity with a small coefficient $\nu$ given by
$\nu=0.05\Delta x^{2}/\Delta t$, where $\Delta x$ is the local spatial
resolution and $\Delta t$ is the time step.
Since the fully ionized plasma in the real corona is highly
conductive, we chose to not use explicit resistivity in the magnetic
induction equation, but magnetic reconnection can still be triggered
through numerical diffusion when a current layer is sufficiently
narrow with thickness close to the grid resolution. By this, we
achieved an effective resistivity as small as we can with given grid
resolution. In the energy equation, the adiabatic index is set to
$\gamma=1$, which thus reduces to an isothermal process.
		
To investigate the role of reconnection and torus instability in
initiating the eruption of a pre-existing MFR, we first construct an
initial magnetic configuration that contains a stable MFR. In
numerical simulations, there are various approaches to building an
MFR. These include the transformation from magnetic arcades to MFR
through slow reconnection driven by flux cancellation
flows~\citep{amariCoronalMassEjection2003a, 2003PhPl...10.1971L,
  2010ApJ...708..314A, 2022ApJ...933..200Z}, the
process of magnetic flux emergence and its associated
dynamic~\citep{2001ApJ...554L.111F, leakeSIMULATIONSEMERGINGMAGNETIC2013,   2023ApJS..264...13L}, the formation of MFR through magnetic helicity condensation driven by
supergranular and granular motions~\citep{2022ApJ...934L...9L}, and
the utilization of semi-analytical models for coronal
MFRs~\citep{1999A&A...351..707T,2014ApJ...790..163T,2018ApJ...852L..21T}. Here
we opted to employ the semi-analytical, near force-free
Titov-Demoulin-modified (TDm) model
\citep{2014ApJ...790..163T,2018ApJ...852L..21T} to construct the
initial magnetic configuration. This choice was made because the
alternative methods as aforementioned are more complex and often
require more time to form an MFR. By utilizing the TDm model, we can
efficiently establish the desired magnetic configuration and proceed
to investigate the role of reconnection and ideal instability in
initiating the eruption.
		
The initial condition of our simulation is established by combining
two methods: the regularized Biot-Savart laws (RBSL)
method~\citep{2018ApJ...852L..21T}, which constructs the TDm
model~\citep{2019ApJ...884L...1G}, and the MHD relaxation
method~\citep{2021FrP.....9..224J}, which allows us to obtain an MHD
equilibrium state from the near force-free field of TDm model. The TDm
model incorporates four independent parameters that govern the
structure of the MFR and the background magnetic field. These
parameters include the path of the axis $\varsigma$, the minor radius,
the magnetic flux $F$, and the electric current of the flux rope. The
path of the axis follows an iso-contour, $B_q = {\rm const}$ (the
background field), which takes the form of a circular arc in the
vertical plane of symmetry of the configuration. This circular arc is
closed by a sub-photospheric arc $\varsigma^{\ast}$, forming a circle
with a radius of $R=75$~Mm. The radius of the torus is set to
$a=20$~Mm. The typical magnetic flux is determined by averaging the
absolute values of the magnetic flux within the radius of the torus
from the two footprints of the flux rope, namely,
$F=(\arrowvert F_+\arrowvert+\arrowvert F_-\arrowvert)$. The typical
electric current is calculated as $I=5\sqrt{2}/(3\mu_0a)$. The
background field $B_q$ is modeled by two points sources
below the bottom boundary. 

\begin{figure*}
  \centering
  \includegraphics[width=\textwidth]{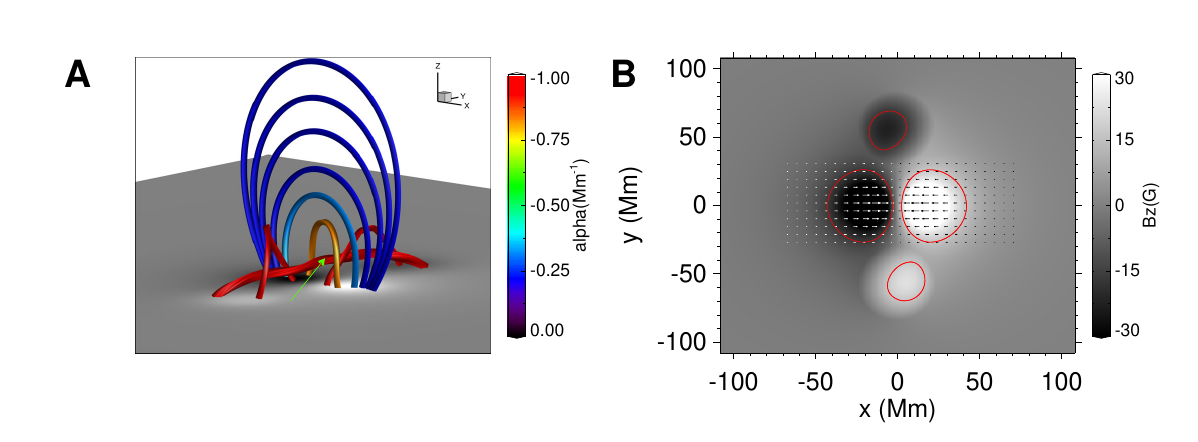}
  \caption{Initial magnetic field (A) and bottom driving flow (B) of the MHD simulation. In panel A, the field lines are shown by the thick lines, which are colored by the value of the nonlinear force-free factor defined as $\alpha = \vec J \cdot \vec B / B^{2}$. The green arrow points out the axis of magnetic flux rope. The bottom surface is shown with the magnetic flux distribution. In panel B, the driving flow at the bottom surface is shown by the arrows, and the background also shows the magnetic flux distribution . The red contour lines indicate the main magnetic polarities.}
  \label{f1}
\end{figure*}
	
Since the TDm magnetic field is not accurately force-free, it is
further relaxed to an MHD equilibrium with a background plasma
atmosphere. The plasma is initially an isothermal gas of temperature
$T = 1 \times 10^{6}$~K in hydrostatic equilibrium and stratified by
the solar gravity with density
$\rho = 2.29 \times 10^{-15}$~g~cm$^{-3}$ at the bottom. The values
for temperature and density are typical in the corona. At the bottom
boundary, the magnetic field evolves according to magnetic induction
equation by treating the bottom boundary as a perfectly line-tying and
fixed surface (with zero velocity) for magnetic field lines. It is
important to note that this does not imply that all magnetic field
components are fixed on the boundary because although the velocity is
zero at the bottom boundary, it may not be zero at neighboring inner
points. Therefore, the horizontal components of magnetic field can
vary with time, until all velocity approaches zero. Figure~\ref{f1}A
shows the relaxed magnetic field configuration of the TDm model, with
colors representing the value of nonlinear force-free factor
$\alpha = \vec J \cdot \vec B/B^2$. This mimics a typical coronal magnetic configuration with a low-lying MFR along the PIL stabilized by the overlying near-potential field,  
and the decay index at the apex of the rope axis
is $0.6$.


With the MHD equilibrium containing a stable MFR configuration as the initial condition, we drove its evolution by introducing a surface converging flow~\citep{2003ApJ...585.1073A} at the bottom boundary
defined as
\begin{equation}
  v_x = -c_0(B_q)_z,  v_y = 0,  v_z = 0
\end{equation}
where $c_0$ is a constant for scaling such that the largest velocity
is $7.4$~km~s$^{-1}$ and $(B_q)_z$ refers to the vertical component of
the background field at the bottom boundary at the initial time. Such
a converging flow can effectively inject magnetic energy into the
system and increase the decay index at the apex of MFR axis by
modifying the flux distribution of the background field, therefore
driving the system to approach an unstable state. Also, the converging motion is favorable for formation of an internal current sheet above the PIL and subsequent triggering of reconnection~\citep{bianRolePhotosphericConverging2022}. Figure~\ref{f1}B
shows the distribution of the magnetic field on the bottom surface,
where the arrows represent the converging velocity.

The simulation volume used in our study is a cube with dimensions of
$[-230, 230]$~Mm in both $x$ and $y$ direction, and $[0, 460]$~Mm in
$z$ direction. This size was chosen to be much larger than the initial
MFR structure, thus allowing sufficient space for simulating the
eruption initiation process of the MFR without the influence from the
side and top boundaries.\footnote{All the simulation runs in this work
  are stopped before the disturbance by the eruption reaches any of
  the side and top boundaries.} The plasma density, temperature, and
velocity were fixed on the side and top boundaries. Furthermore, the
horizontal components of the magnetic field were linearly extrapolated
from the inner points, while the normal component was modified to
satisfy the divergence-free condition, thus preventing any numerical
magnetic divergence from accumulating at the boundaries.

The entire simulation volume is resolved using a block-structured AMR
grid. We have carried out three set of simulations (to be referred to
as RES0, RES1, and RES2), in which the base resolution is identical
with $\Delta x = \Delta y = \Delta z = \Delta = 2.88$~Mm and the
highest resolution is, respectively, $\Delta =180$~km, $90$~km, and
$45$~km. The highest resolution in the simulations is used to resolve
the core region of the computational volume with strong magnetic field
and current density (for instance, the region encompassing the MFR), and also to dynamically capture thin layers of
enhanced current density that may develop into current sheet. By using
the different-resolution experiments, we can test whether our
calculation converges in the ideal MHD regime. A more important
consideration for performing the experiments on different resolutions
is that, with all other settings being the same, in the simulation of
a higher resolution a current sheet can be thinner and thus the onset
of reconnection will be later than that of a lower resolution. This is
because, as we have mentioned before, the reconnection results from
numerical diffusion when the thickness of the current sheet is close
to the grid resolution. In this sense, the onset time of reconnection
depends on the grid spacing, because, with a smaller grid size, a
thinner current sheet can develop and sustain a stronger current
density (and therefore more free energy), which needs more time to
accumulate. Therefore, the onset of reconnection in the current sheet
is postponed relative to runs with lower resolutions. The dependence
of reconnection onset time and the effective resistivity on the grid
resolution in our MHD code has been carefully studied in previous
simulations~\citep{2021NatAs...5.1126J}.
		
\begin{figure*}
  \centering
  \includegraphics[width=\textwidth]{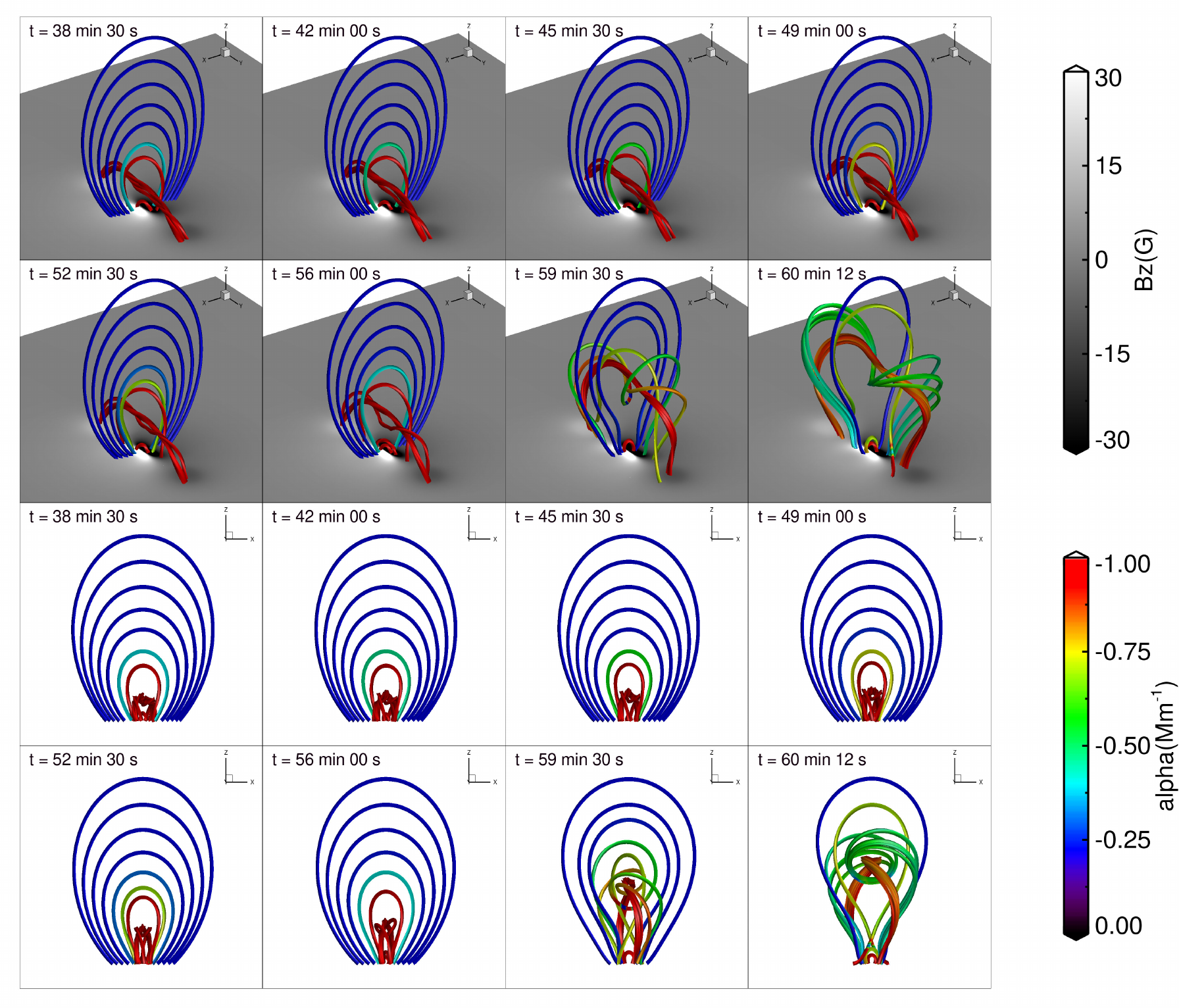}
  \caption{Evolution of magnetic field lines in the simulation. (A) The 3D prospective
    view of magnetic field lines. The colored thick lines represent magnetic field line and the colors denote the value of $\alpha = \vec J \cdot \vec B / B^{2}$, which
    indicates how much the field lines are non-potential. (B) side view
    of the same field lines shown in (A). An animation is provided.}
    \label{f2}
\end{figure*}

\begin{figure*}
  \centering	
  \includegraphics[width=\textwidth]{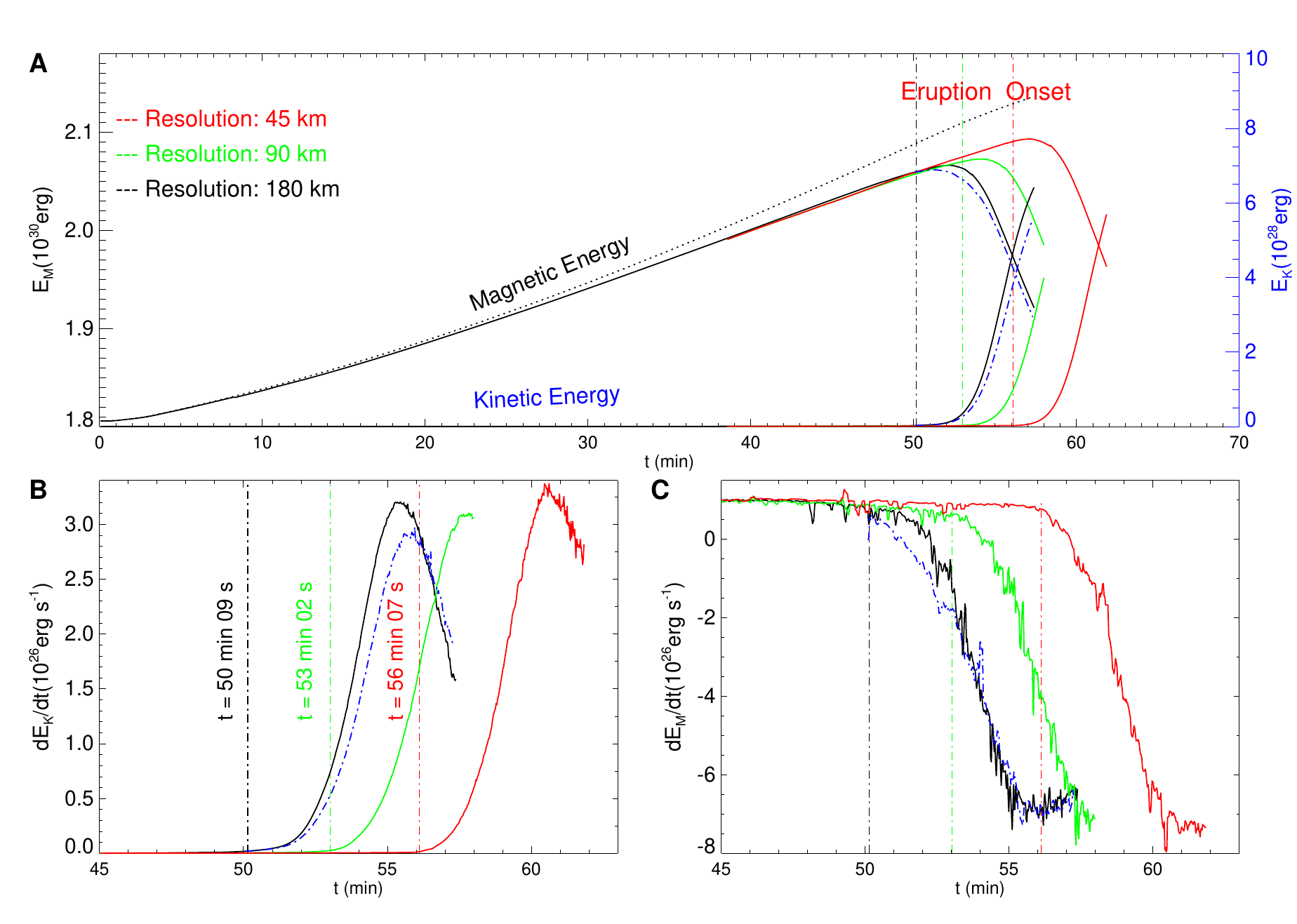}
  \centering
  \caption{Evolution of different parameters of energies in the simulation runs with different grid resolutions.
   (A) Magnetic energy $E_M$ and kinetic energy $E_K$. The colors of the curves denote the results of the different runs; black for RES0, green for RES1, and red for RES2. The dotted curve shows the energy injected into the volume (i.e., time integration of total Poynting flux) from the bottom boundary through the surface flow. The blue dot-dashed curves show the result for RES0 with the bottom driving flow stopped once the eruption is triggered. (B) Increasing rate of kinetic energy $dE_K/dt$. (C) Releasing rate of magnetic energy $dE_M/dt$. In all panels, the vertical dot-dashed lines mark the eruption onset times, which are different in the different runs, as denoted by the numbers in panel B. }
    \label{f3}
\end{figure*}

\section{Results}
\label{results}
		
As driven by the converging motion, the coronal magnetic field first
undergoes a quasi-static evolution and then runs into a drastic
eruption, as depicted by the 3D magnetic field lines in
Figure~\ref{f2} (and the animation) and the energy curves in Figure~\ref{f3}. In the
quasi-static evolution stage, the overlying field expands slowly which
allows a continuous rise of the MFR but the system maintains in
proximity to a force-free state. The magnetic energy increases almost
linearly, while the kinetic energy remains negligible as compared with
the magnetic energy. This suggests that almost all the energy as
injected into the corona by the surface flow through the Poynting flux
is stored in the magnetic field. At a critical point, i.e., onset of
the eruption, the MFR starts to rise rapidly along with reconnection
in its wake. During the eruption, the kinetic energy increases
impulsively by nearly two orders of magnitude in a very short period
of around 5~min, indicating the rapid acceleration of the
plasma. Despite the continual injection of Poynting flux through the
bottom surface, the magnetic energy drops immediately, indicating a
quick release of magnetic energy during the eruption. From the energy
curves, it can be seen that the eruption starts at different moments
in the three simulation runs with incremental higher resolution from
RES0 to RES2, which are $t=50$~min~09~s, 53~min~02~s, and 56~min~07~s,
respectively. We have also carried out a test run of RES0 by stopping
the surface driving motion once the eruption begins (see the dashed
curves in Figure~\ref{f3}), and validated that the eruption, once
triggered, is almost not affected by the slow surface driving motion.

\begin{figure*}
  \centering
  \includegraphics[width=\textwidth]{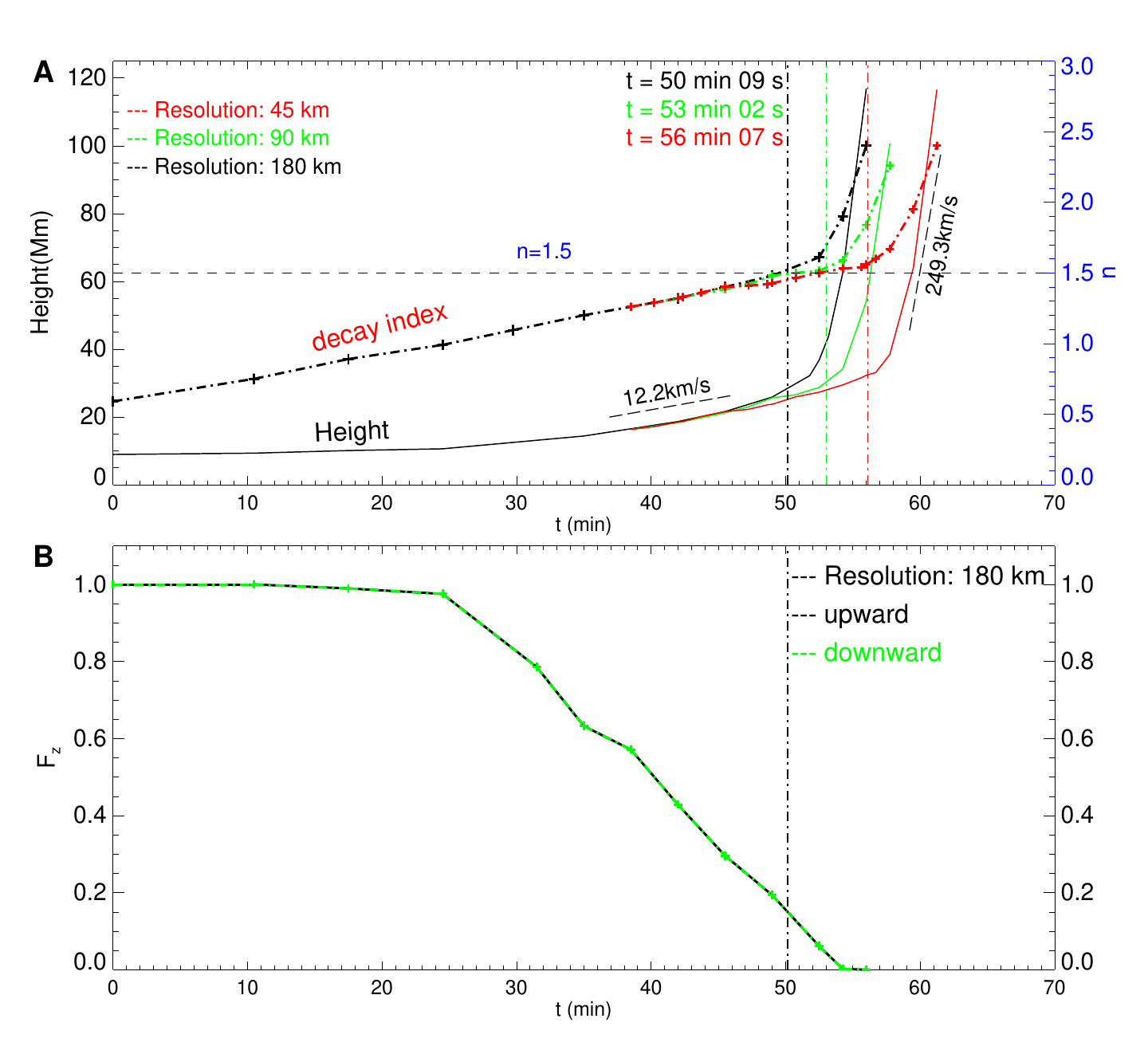}
  \caption{Evolution of different parameters of the MFR in the simulation. (A) Apex height (solid lines) of the MFR axis and the decay index (dashed lines) of the strapping field at the apex of the MFR axis. Colors of the lines are used to denote the results in the different runs from RES0 to RES2. The horizontal dashed line denotes the critical height for decay index $n=1.5$. The vertical dashed dotted lines denote the onset times of the eruptions. The average speed of the MFR axis in the slow rise phase and during the eruption is shown with the oblique dashed lines. (B) The upward (colored in black) and downward (green) components of Lorentz force at the apex of the MFR axis. They are defined respectively as $F_z^{\rm upward} = (\vec J \times \vec B_{\rm MFR})_z$, and $F_z^{\rm downward} = (\vec J \times \vec B_{e})_z$ (where $\vec B_{\rm MFR}$ is the magnetic field of the MFR and $\vec B_{e}$ the strapping field). Result is shown for RES0 run. Note that the two plots overlay each other almost exactly, indicating that the upward and downward forces are nearly exactly balanced over the entire period.}
\label{f4}
\end{figure*}

\begin{figure*}
  \centering
  \includegraphics[width=\textwidth]{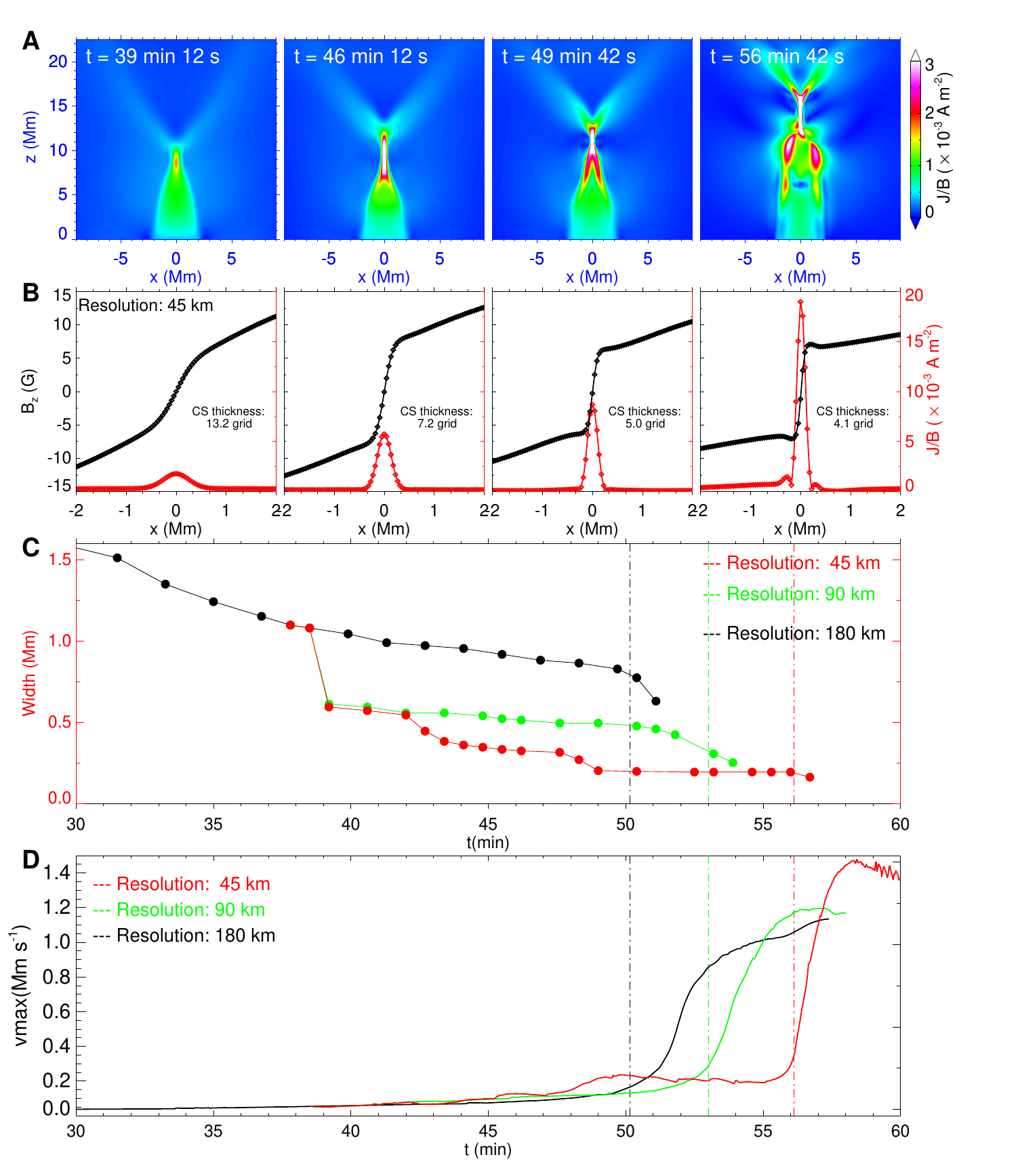}
  \caption{Formation of current sheet and onset of reconnection. (A)
    Distribution of current density $J$ normalized by magnetic field
    strength $B$ on the vertical cross section (the $y = 0$
    slice). (B) 1D profile of the magnetic field component $B_z$ and
    $J/B$ along a horizontal line crossing perpendicular to center of the current sheet
    (which is defined as the point with largest $J/B$). The diamonds denote
    values on the grid nodes. (C) Thickness evolution of the current sheet before the eruption onset. The results are shown for all the three runs from RES0 to RES2. (D) The largest velocity in the three different runs. \label{f5}. The vertical dashed lines show  the onset time of eruptions in the different runs.}
\end{figure*}

Figure~\ref{f4}A depicts the time evolution of the apex height of the
MFR axis and also the corresponding decay index of the strapping field
there. The simulations all show that the height curves of the MFR axis
exhibit a gradual growth before the eruption, and then rises rapidly with
onset of the eruption. The slow rise of the MFR with a speed of around $10$~km~s$^{-1}$ in 20 minutes before the eruption onset is consistent with many observations that shows filaments often rise slowly before they erupt more quickly~\citep{zhangTemporalRelationshipCoronal2001,
 sterlingSlowRiseFast2005, mccauleyProminenceFilamentEruptions2015}.
 Prior to the eruption the decay index grows
almost linearly, even in the early phase (for example, from $t=0$ to
$30$~min) when the amount of rise of the MFR axis is very small. This
is because the decay index is changed in a larger portion due to variation
of the strapping field (as driven by the converging flow) than the rise of the MFR. Importantly, the decay index increases
continuously from the initial value of $0.6$ to a value very close to
the canonical threshold $n_c = 1.5$ at which the eruption immediately
begins. This good consistence of the eruption onset with the threshold
condition of the torus instability appears to indicate that the
eruption of the MFR is triggered by torus instability. Nevertheless,
whether the torus instability can actually occur also
depends on the details of the MFR itself rather than solely the decay index of the external field. Correspondingly, the critical decay index for torus instability has a wide range of around $1.1 \sim 1.7$ rather than the specific number of $1.5$, which has been mentioned in Section~\ref{sec:intro}. More intricately, in some case, for example as shown in a failed eruption model recently developed by~\citet{jiangModelFailedSolar2023}, an MFR erupts with the decay index at its axis being as small as $1$, while the eruption is halted even when the decay index at the MFR axis has reached around $2.5$! Therefore it is still unclear whether the torus instability sets in by only calculating the decay index. To look into whether there
is really a loss of force balance at the axis of the MFR at the eruption onset time, which is predicted by the torus instability,
we calculated the upward and downward forces at the axis. They are defined respectively as
$F_z^{\rm upward} = (\vec J \times \vec B_{\rm MFR})_z$,
and $F_z^{\rm downward} = (\vec J \times \vec B_{e})_z$ (where $\vec B_{\rm MFR}$ is the magnetic field of the MFR and $\vec B_{e}$ the strapping field).
The evolution of these two forces is shown in \Fig~\ref{f4}B. Both the two forces decrease with time and thus with height since the axis rises continuously. However, these forces are well balanced, even at the eruption onset time, suggesting that the torus instability is not actually triggered.

\begin{figure*}
  \centering
  \includegraphics[width=0.7\textwidth]{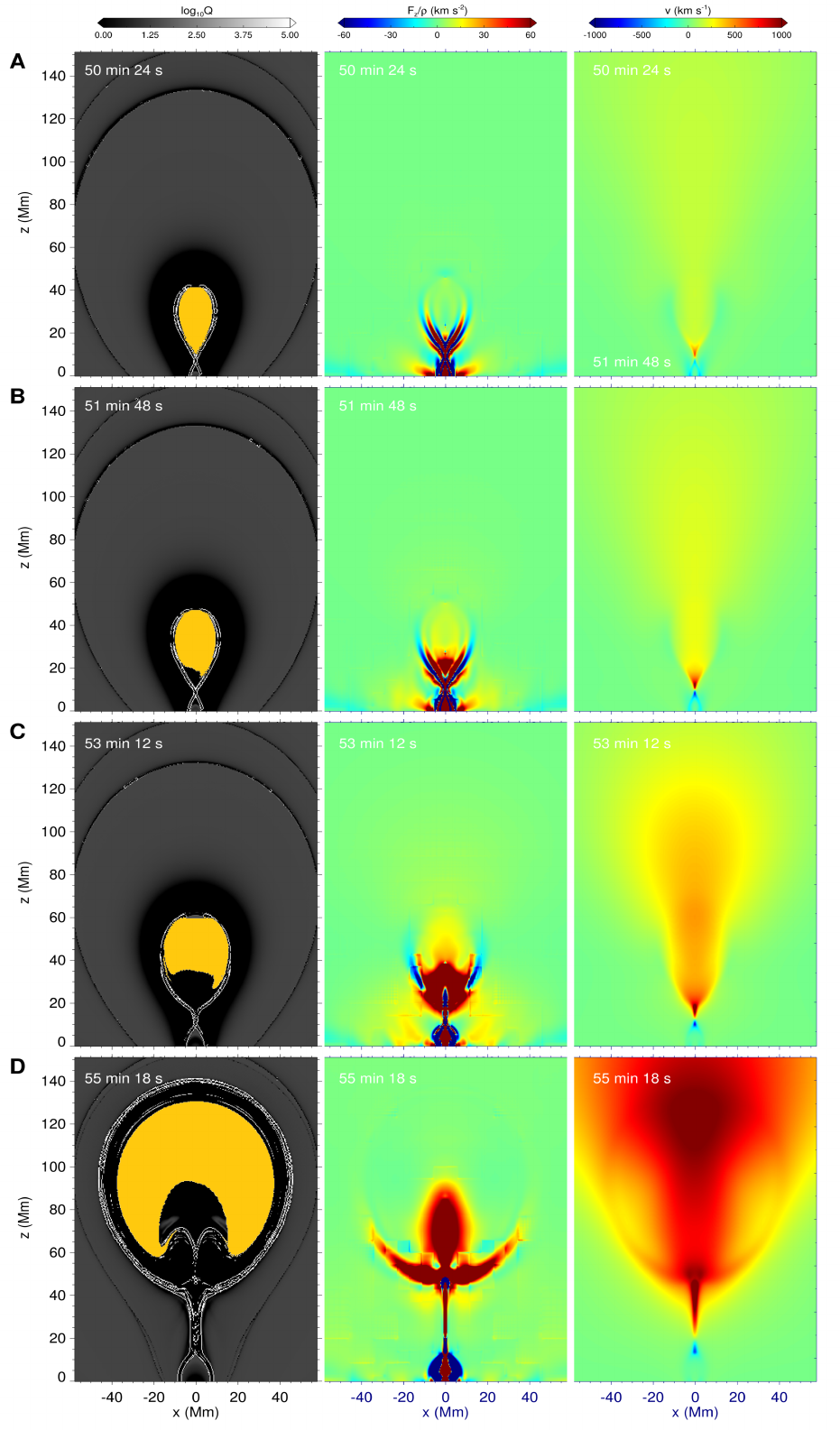}
  \caption{Eruption of the MFR as shown at the central vertical cross section $y=0$ slice. Left: The magnetic squashing factor $Q$. The areas colored in yellow show the region as occupied by the magnetic flux belonging to pre-eruption MFR, i.e., the original component of the erupting MFR. Middle:  The upward component of Lorentz force $F_z$ as divided by density $\rho$. Right: The vertical velocity $v_z$. Time proceeds from top to bottom: the first moment is close to the onset time of the eruption, and the last moment is close to the peak time of the eruption acceleration. }
  \label{f6}
\end{figure*}

On the other hand, the simulations indicate the importance of
reconnection in triggering the eruption. As shown in \Fig~\ref{f5}, a
thin current layer is formed below the MFR during the quasi-static
evolution stage, as driven by the slow converging motion at the bottom
surface. The current layer gradually becomes thinner, especially in
its core region where the current density is the
highest. \Fig~\ref{f5}C shows the time variation of the thickness of
the current layer. The thinning of the current layer occurs
quasi-statically. Eventually, the current layer evolves into a current
sheet as it thins down to a spacing of around 3 grids, leading to reconnection
in the current sheet. The magnetic field component $B_z$ crossing the
current sheet shows a thinning down to a tangential discontinuity in
numerical sense. The moment at which the current sheet forms and
reconnection sets in is exactly the same as the eruption onset time in
the different resolution runs. As can be seen in \Fig~\ref{f5}D, the
start of reconnection in the different runs is also indicated by the
sudden rise of the maximal velocity $v_{\rm max}$ in the computational
volume. This is because the sharp variation of the maximal velocity
occurs in the reconnection outflow region, where the plasma is
impulsively accelerated upward (see also the right column of \Fig~\ref{f6}).

Both the height profile of the MFR axis (Figure~\ref{f4}) and the
energy evolutions (Figure~\ref{f3}) show that the eruption onset is
sensitive to the resolution, as it is postponed by about 3~min
incrementally from RES0 to RES3. This is because, as has been noted
near the end of Section~\ref{model}, at a higher resolution a current
sheet can be thinner and the onset time of reconnection is postponed
(see Figure~\ref{f5}C). Our analysis indicates that the reconnection
of the core current sheet plays a more important role in determining
the onset of the eruption (and fast rise of the MFR) than the torus
instability. Otherwise, an eruption as initiated totally by an ideal
MHD process will not be affected by the grid resolution since the here
the lowest RES0 is already high enough to resolve the ideal MHD
process.

To further investigate the source of driving force for the
acceleration of the eruption, we divide the flux (and thus the volume) of entire erupting
MFR into two components and calculate the works done by these two
components separately, for the aim of distinguishing the roles of the
ideal instability and the reconnection played in driving the eruption. Specifically, one component,
referred to as the ``original component'', is the magnetic flux
originated from the pre-existing MFR before the eruption. This
component expands in volume but remains to be the same magnetic flux
during the eruption, and if the MFR evolves ideally without
reconnection, the eruption will be driven by only this
component. Therefore, the work by Lorentz force in the original
component corresponds to contribution of the ideal instability. The
other component is the flux newly incorporated into the MFR through
the continuous reconnection during the eruption, which will be called
the ``reconnected component'', and the work associated with this
component should be attributed to the reconnection, since this work
does not exist without reconnection. The reconnected component wraps
the original component and both expand in volume during the eruption.
		
We calculated the magnetic squashing degree ($Q$ factor) to identify
the quasi-separatrix layers (QSLs) ~\citep{2016ApJ...818..148L} that
define the 3D boundary of the entire erupting MFR, for which the
volume is denoted as $V_e(t)$. In practice, it does not need to
compute the $Q$ factor in the full 3D volume (which is extremely
time-consuming) but only on the central cross section of the volume
(i.e., the $y=0$ slice in our case). The QSLs on the slice define a closed area
separating the MFR from the ambient field, which are shown in the left
column of Figure~\ref{f6}. We traced all magnetic field lines that
pass through the grid points within the closed area. Each of these
field lines defines an elementary flux tube that constitutes the MFR, and
$V_e(t)$ is the sum of the volumes of all these flux tubes. Since the
axial flux of the elementary flux tube is conserved along the tube,
the cross section of the flux tube at different positions is
inverse to its local magnetic field strength, and thereby the volume of
the flux tube can be obtained easily. The original component is
obtained by tracing the field lines (or the elementary flux tubes)
that root in the same footpoints of the MFR at the eruption onset
time, and the volume is denoted as $V_o(t)$. Then the volume occupied
by the reconnected component, $V_r(t)$ is obtained by subtracting the
original component from the entire MFR, $V_r(t) = V_e(t) - V_o(t)$. In
the left column of Figure~\ref{f6}, the original component at different times is
denoted by the yellow area on the $y=0$ slice. With these volumes
determined, we can calculate the powers and works done by the upward
Lorentz force $F_z = (\vec J \times \vec B)_z$ in these two components
of the MFR using the following formulas,
\begin{eqnarray}
  P_c(t) = \int_{V_c, F_z>0, v_z > 0} F_{z} v_z {\rm d}V, \nonumber \\
  W_c (t) 
    = \int_{t_{\rm onset}}^{t} P_c(t) {\rm d}t,
\end{eqnarray}
where the subscript $c$ refers to $o$ (by the original component), $r$ (reconnection component), or $e$ (entire). By restricting the
computation within domain of $F_z > 0$ and $v_z > 0$, the power and work
as obtained account for the upward acceleration of the eruption.
		
\begin{figure*}
  \centering
  \includegraphics[width=\textwidth]{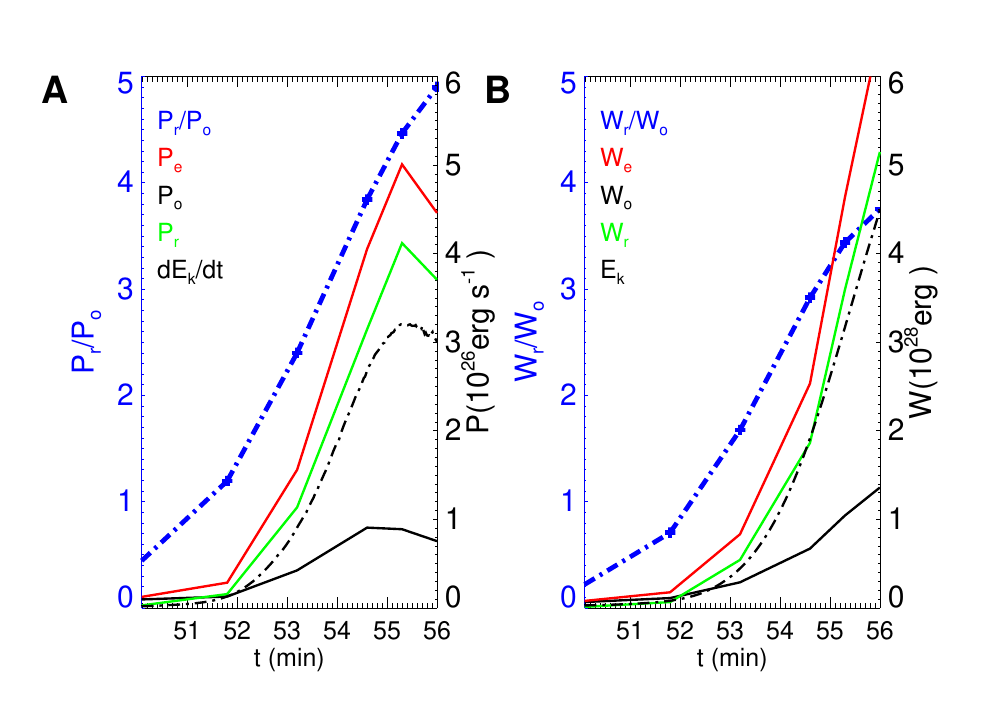}
  \caption{Evolution of the powers (A) and works (B) done by the Lorentz force of the erupting MFR that drives the upward acceleration of the eruption. The time starts from the onset of the eruption $t=50$~min~$09$~s. The red solid lines represent the power and work done by the entire MFR, $P_e$ and $W_e$. The black solid lines represent the those done by original component of the MFR, $P_o$ and $W_o$. The green solid lines represents those as contributed by the newly reconnected component of the erupting MFR, $P_r$ and $W_r$. The black dashed line in A are shown for increasing rate of the kinetic energy, and in B for evolution of kinetic energy. The blue dashed lines represent the ratio of powers (A) and works (B) from the reconnected component to those from the original component.}
  \label{f7}
\end{figure*}

The results are shown in Figure~\ref{f7} for the simulation run RES0
(other runs have almost the same results and are thus not shown) with
time starting from the onset of the eruption. In comparison, the
evolution of kinetic energy is also shown in the figure. With start of
the eruption, the powers ($P_e$, $P_o$, and $P_r$) first increase to
their peak values at around $t=55$~min and then decrease
(Figure~\ref{f7}A). This is consistent with time profile of the
kinetic energy increasing rate, which also reaches its peak at nearly
$t=55$~min. At the beginning, $P_o$ (the black solid line) is large
than $P_r$ (the green solid line), but is quickly overtaken by the
latter after about $1$~min. This is clearly seen from the ratio of
$P_r/P_o$, which increases to a high value of $5$ near the peak time,
showing that the reconnected component dominates the acceleration of
the eruption at the peak time.  As seen in Figure~\ref{f7}B, all the
works continuously increase, consistent with the profile of the
kinetic energy.
Also same as the profiles of the power, the work done by the
reconnected component $W_r$ increases much faster than the original
component $W_o$, and exceeds well $W_o$ after around $t=52$~min. At
$t=56$~min, $W_r$ is $3.8$ times of $W_o$ (see the ratio $W_r/W_o$),
and thus it accounts for over $79\%$ of the total work. The dominant
role of the reconnected flux in driving the eruption can be
appreciated by inspecting the distribution of the Lorentz force in the
erupting MFR, which is shown in the middle column of \Fig~\ref{f6}. The upward
acceleration by the Lorentz force $F_z/\rho$ is mainly distributed in
the lower part of the erupting MFR, i.e., in the newly reconnected
flux that joins in the MFR from the reconnection site, while the
acceleration at the original component of MFR is much
weaker. Therefore the original component is mainly pushed upward from
below by the reconnected component. Consequently, the lower part of the original component (the yellow area in the left column of \Fig~\ref{f6}) is compressed upward substantially, and its boundary is transformed from an upward concave shape to a downward concave one.

This analysis indicates that the work done by
reconnected component plays a major role in driving the eruption,
while the original component only contributes a small part to the
driving force of the eruption. In other words, reconnection plays a
more important role in driving the eruption, while torus instability
is only a minor contributor to the driving force of the eruption.

\section{Conclusions}
\label{concl}
		
To summarize, we have developed a numerical model of solar eruption of
a pre-existing coronal MFR in which both the ideal and resistive
processes contribute to the initiation process but the latter plays
the primary role. Unlike many other models that start from an unstable
MFR configuration and use some kind of disturbance to trigger the
eruption, we first constructed a stable MFR enclosed by a background
field by relaxing a TDm MFR (utilizing the RBSL method) to an MHD
equilibrium. This initial state was then driven to evolve into an
eruption by a continuous converging motion of the footpoints of
background field at the bottom surface. Therefore our model can show
the self-consistent transition from a stable equilibrium to an
eruption as driven by photospheric motions. The coronal magnetic field
underwent first a quasi-static energy storage phase, in which magnetic
energy is gradually injected through the Poynting flux as introduced
by the converging flow. In this quasi-static evolution phase, the MFR
is driven to rise slowly with the decay index at the apex of rope axis
increases continuously. Meanwhile, a thin current layer is formed
beneath the MFR, as a result of the rise of the MFR and the converging
of the background arcade.

Interestingly, the eruption is initiated at a critical point when two
conditions are met almost simultaneously. Firstly, the MFR axis
reaches a height where the decay index reaches very close to
$n_{c} = 1.5$, which is consistent with the canonical threshold of the
torus instability. Secondly the current layer is thinned to a current
sheet and reconnection sets in. The first condition seems to indicate
that the torus instability triggers the eruption and the decay index
plays a crucial role in determining the occurrence of eruptions,
while the second condition appears to infer that the eruption is
actually triggered by reconnection. To clarify this issue, we have analyzed the profiles of the hoop force
and strapping force at the MFR axis, and found that these forces are well balanced at the eruption onset time, suggesting that the eruption is not triggered by torus instability. Additionally, we have performed experiments with different grid resolutions and
found that the eruption is postponed with higher resolutions, which
highlights the significance of reconnection in determining the timing
of eruption, because at a higher resolution a current sheet is thinner
and the onset time of reconnection is postponed. We further analyzed
the driving force for the acceleration of the eruption by quantifying
separately the work by the original MFR (which exists before the
eruption) and that by the newly reconnected magnetic flux that joins
in the erupting MFR. It is found that the work done by the reconnected
component accounts for a major part (about $80\%$) in the total work
done by the MFR during the eruption. This stresses the importance in
of reconnection in driving the eruption. Therefore our study suggests
that, even in the situation with a pre-existing coronal MFR,
reconnection could play a primary role in triggering and driving its
eruption.

The key scenario as demonstrated in this paper, that is, a current
sheet is formed during the quasi-static evolution phase and eruption
is mainly initiated by the reconnection at the current sheet, is
essentially identical to the fundamental mechanism as shown in recent
high-resolution simulations~\citep{2021NatAs...5.1126J, Bian2022a,
  Bian2022b} for a single sheared arcade (without pre-existing
MFR). Therefore our study shows that the fundamental mechanism can
also be extended to the situation even with a pre-existing MFR. This
study also suggests that the pre-existence of MFR, as often found
before eruption, does not necessarily indicate the ideal instabilities
as being responsible for the eruption. This is consistent with recent observational studies. For example, by a survey of the
major flares (above GOES M5.0) from 2011 to 2017 with coronal magnetic
field extrapolations, \citet{duanStudyPreflareSolar2019} shows that
nearly $90\%$ of the events possess pre-flare MFRs, which indicates
that the pre-existence of MFR is rather common for major
flares. However, there are over half of MFR-possessing events cannot
be explained by the MFR instabilities, since for these events the
controlling parameters (decay index and twist degree) of the MFR
before flare are apparently below their thresholds for triggering the
MFR instabilities. \citet{duanStudyPreflareSolar2019} concluded that
these events might be triggered by magnetic reconnection rather than
MHD instabilities, and the scenario as shown in this
study may apply to these events. Moreover, cautions should be taken when interpreting the
eruption initiation by the torus instability using the decay index,
since our study suggests that although the decay index is close to the
threshold, the eruption only begins with onset of the reconnection.

Nevertheless, it should be noted that our model is much more
simplified if compared to the realistic events, which has a higher
degree of complexity in either the coronal configuration and the
photospheric motions~\citep{jiangStudyComplexMagnetic2022}. Future studies will be performed using
data-constrained and data-driven
simulations~\citep{jiangDatadrivenModelingSolar2022} for real events
to disclose the key mechanism in triggering and driving solar
eruptions.

\section*{Acknowledgements}
This work is jointly supported by National Natural Science
Foundation of China (NSFC 42174200), Shenzhen Science and Technology
Program (Grant No. RCJC20210609104422048), Shenzhen Technology Project
JCYJ20190806142609035, Shenzhen Key Laboratory Launching Project
(No. ZDSYS20210702140800001), Guangdong Basic and Applied Basic
Research Foundation (2023B1515040021) and the Fundamental Research Funds for the Central Universities (Grant No. HIT.OCEF.2023047). The computational work was
carried out on TianHe-1(A), National Supercomputer Center in Tianjin,
China. 

\section*{Data Availability}
All the data generated for this paper are available from the authors
upon request.











\bsp	
\label{lastpage}
\end{document}